\documentclass[a4paper]{jpconf}
\usepackage{graphicx}
\begin{document}
\title{Open questions with ultra-high energy cosmic rays}

\author{Pasquale Blasi}

\address{INAF/Osservatorio Astrofisico di Arcetri, Firenze, Italy}

\ead{blasi@arcetri.astro.it}

\begin{abstract}
We briefly discuss three aspects related to the origin of
ultra-high energy cosmic rays (UHECRs) namely: 1) particle
acceleration in astrophysical sources; 2) transition to an
extragalactic origin; 3) spectrum and anisotropies at the highest
energies. 
\end{abstract}

\section{Introduction}

Astrophysical acceleration processes for cosmic rays need to be 
pushed to the extreme
in order to account for energies as high as the observed ones, of the
order of $\sim 10^{20}$ eV. This does not imply that no class of
sources can explain what we observe. On the contrary the detection of
such high energy particles is a unique opportunity to learn about how
particle acceleration works in extreme conditions. On the
other hand, even if a class of sources is found that are, at least in
principle, able to accelerate cosmic rays up to energies in excess of
$10^{20}$ eV, the attenuation of the flux due to the process of
photopion production induces a flux suppression at $\sim 10^{20}$ eV, 
the so-called GZK feature \cite{gzk1,gzk2}. The details of such 
suppression
depend however on numerous astrophysical factors (e.g. injection
spectrum, intrinsic maximum energies at the accelerators, redshift
evolution of the source luminosity, spatial inhomogeneity in the
source distribution and strength and topology of the intergalactic
magnetic field).   

In order to understand the origin of extragalactic cosmic rays and more
specifically of UHECRs, it is crucial to understand which cosmic
rays are in fact extragalactic. The two main lines of thought in
this respect will be summarized and discussed: in the so-called {\it
ankle scenario} \cite{wibig,hillas} the
transition takes place around $10^{19}$ eV where a steep galactic
spectrum encounters the flat spectrum of extragalactic cosmic rays.
In the {\it dip scenario} \cite{dip1,dip2} the
transition takes place at energies roughly one order of magnitude
lower. The case of a mixed chemical composition of extragalactic
cosmic rays \cite{par1,par2} will also be discussed.

The paper is structured as follows: in \S \ref{sec:acceleration} we
discuss the problem of acceleration and some recent findings on
magnetic field amplification in the acceleration region. In \S
\ref{sec:trans} we discuss the current ideas on the transition
between galactic and extragalactic cosmic rays. In \S \ref{sec:GZK}
we describe the current status of the observations of the end
of the cosmic ray spectrum and of the anisotropies in such energy
range. We conclude in \S \ref{sec:concl}.

\section{How does Cosmic Ray Acceleration work?}
\label{sec:acceleration}

In this section we illustrate some recent developments of studies of 
particle acceleration at shock waves that may have important
consequences for our understanding of the origin of UHECRs, although
at the present stage most of these investigations are aimed at
supernova remnants (SNRs), the possible sources of lower energy
galactic cosmic rays.  
 
In the case of SNRs, acceleration in assumed to take place at the
shock front associated with the supersonic motion of the expanding
shell. Particles are energized through diffusive acceleration {\it a
la Fermi}. 

It is well known that the mechanism of diffusive particle acceleration 
at supernova shocks is efficient only if the level of scattering in
the shock vicinity is much larger than that warranted by the
interstellar medium turbulence. This condition may be fulfilled 
either because the circumstellar material provides the scattering, or
because the accelerated particles generate a larger magnetic field
$\delta B$ from the background field $B$, through streaming
instability. The possibility of magnetic field amplification was
already discussed in Refs. \cite{bell78,lagage83a,lagage83b}, where
the maximum achievable energy was evaluated. The conclusion of 
\cite{lagage83a,lagage83b} was that shocks in SNRs could accelerate 
cosmic rays up to $\sim 10^4$ GeV, below the knee, if the
amplification results in $\delta B/B\sim 1$ (on all spatial scales) 
and the diffusion coefficient has the form of the Bohm diffusion.

The maximum value of the amplified magnetic field $\delta B$ is
however not limited by the value of the background field, but rather
by $\delta B = B \left[ 2 M_A \frac{P_{CR}}{\rho u^2} \right]^{1/2}$,
where $\rho u^2$ is the ram pressure of the inflowing fluid in the
upstream region, $P_{CR}$ is the pressure in the form of accelerated
particles, and $M_A = u\sqrt{4\pi \rho}/B$ is the Alfvenic Mach
number of the upstream fluid. All these results are obtained in the context
of the quasi-linear theory and should in principle be used only
for $\delta B/B\ll 1$, while they predict $\delta B/B\gg 1$,
therefore these conclusions should be taken with much care. The waves
that turn nonlinear within this approach are Alfv\'en waves.

Recently the authors of Ref. \cite{lb00} have presented an approach
that would lead to magnetic fields much larger than those discussed
above. In \cite{bell2004} the appearance of a new non-alfvenic purely  
growing mode was discussed and was found to saturate at
\begin{equation}
\delta B = B M_A \sqrt{\frac{u}{c}\frac{P_{CR}}{\rho u^2}},
\end{equation}
where the symbols have the same meaning as above. This saturation
level, determined in the context of quasi-linear limit, seems to 
be confirmed by hybrid simulations \cite{bell2004}. 
For typical parameters of a SNR and assuming that an
appreciable fraction of the kinetic pressure is transformed into
cosmic rays ($\rho u^2 \approx P_{CR}$), one can predict $\delta
B/B\sim 500$ (versus $\delta B/B\sim 20$ in the previous case). If
one assumes Bohm diffusion, this translates to 
higher values of the maximum energy of accelerated particles: for
$\delta B/B\sim 500$ one has $E_{max}\sim (0.5-2)\times 10^7$ GeV
(assuming Bohm diffusion).

The condition $\rho u^2 \approx P_{CR}$ is typically found to be a
consequence of the dynamical reaction of the accelerated
particles (see \cite{malkov} for a review of non-linear diffusive 
particle acceleration). This reaction, which leads to the so-called
{\it cosmic ray modification} of shocks, has three important
phenomenological consequences: 1) creation of a cosmic ray precursor
which is also responsible for a concave energy spectrum (non power
law); 2) large efficiency for particle acceleration; 3) suppression
of the plasma heating in cosmic ray modified shocks.

A unified picture of nonlinear particle acceleration at shocks with
self-generation of scattering has recently been presented in
\cite{amato1,amato2,don}. An important effect of the shock modification
is that while the amplification of the magnetic field leads to an
increase of the maximum achievable energy, the precursor (slowing
down of the upstream fluid and spatial variation of the magnetic
field) leads to a somewhat lower value of the maximum energy
\cite{blasi2006}. 

It is worth stressing that the KASCADE data (see \cite{hora} for a 
review) show that the proton spectrum extends to $\sim 10^7 GeV$. The 
spectrum of helium nuclei appears to extend to slightly higher
energies, as it could be expected in a rigidity dependent model of
acceleration. In this picture the knee in the iron component would
be expected to be at energy $E_k^{Fe} = Z E_k^p \approx 8\times
10^{16}$ eV, while the spectrum would probably extend up to
$E_{max}^{Fe} = Z E_{max}^p \approx 2\times 10^{17}$ eV. The
spectrum of iron nuclei is however not observed in a reliable way at
the present time, therefore this should be considered as a phenomenological
conclusion, which however, combined with theoretical insights,
hints to the fact that the galactic component of cosmic rays should
end around $\sim 2\times 10^{17}$ eV.

Independent evidences for the magnetic field amplification discussed
above comes from Chandra X-ray observations of the X-ray rim of
several SNRs, resulting from synchrotron emission of
relativistic electrons accelerated at the shock front. It has been
pointed out that the spatial extension of these regions is
compatible with magnetic fields of the order of $\sim 100-300\mu G$
and not with the typical fields in the interstellar medium \cite{warren}.

After learning all these pieces of information from studying the
accelerators of galactic cosmic rays it would be desirable to use them
for a better understanding of the sources and acceleration processes
involved in the production of extra-galactic cosmic rays, most of
which involve relativistic shocks. In order to reach this goal two 
important steps need to be pursued: 1) a mathematical description 
of shock modification for relativistic shocks and 2) a theory of 
magnetic field amplification that applies to plasmas in relativistic 
motion. It is worth keeping in mind that this aspects of particle
acceleration are not just corrections to a well established picture,
but the basic reasons why the process works, at least in galactic
sources.   
In addition to these purely theoretical problems, it is crucial that
the search for the sources of the highest energy cosmic rays take
place in the context of multifrequency and multimessenger astronomy. 

\section{Where does the transition from galactic to extragalactic
  cosmic rays occur?}
\label{sec:trans}

As pointed out in Sect. 2, there are phenomenological arguments that
suggest that the galactic component of cosmic rays may end around
$\sim 2\times 10^{17}$ eV. This conclusion, which still needs to be
confirmed by solid measurements, would lead to the following
conclusions: 1) the extragalactic component starts around this same
energy and 2) the ankle, traditionally interpreted as a feature that
results from the intersection of the two components, would require an
alternative explanation.  

A dip appears in the spectrum of extragalactic cosmic rays at energy
$\sim 3\times 10^{18}$ eV \cite{dip1,dip2,usall} (the position of the
ankle), as due to the combination of adiabatic losses (expansion of the
universe) and $e^\pm$ pair production. When calibrated by the position
of the dip, the spectra of UHECRs determined by all experiments agree
very well with each other.

The low energy part of the dip fits what is currently named the 
{\it second knee}. Below the second knee the predicted spectrum
flattens and drops below the flux of galactic cosmic rays. This low
energy part depends to some extent on the mean distance between
sources and on the magnetic field value and topology in the
intergalactic medium \cite{alo,lemoine}. In
this scenario the transition between galactic and extragalactic
cosmic rays takes place somewhere between $10^{17}$ eV and $10^{18}$
eV, but it remains true that in the transition region a steep
galactic spectrum encounters a flatter extragalactic spectrum
\cite{usall}, as in the {\it ankle} scenario.

Both possibilities are currently viable and have positive and
negative aspects. The injection spectrum required to fit the data in
the ankle scenario is as flat as $E^{-\alpha}$, with $\alpha\sim
2-2.4$, which is tantalizingly close to the results expected for
shock acceleration, while in the dip scenario the injection spectrum
has $\alpha=2.6-2.7$. However, it was pointed out in
Ref. \cite{kache,usall} 
that the superposition of flat spectra with different
maximum energies naturally provides a good fit to the data without
requiring a steep injection spectrum. In any case one should always
keep in mind that several pieces of physics lead to a steepening of
the spectra with respect to the canonical case $\alpha\sim 2$ 
\cite{usall}.

From the point of view of the chemical composition the two models
differ the most: in the ankle scenario \cite{wibig}, 
galactic cosmic rays extend to $>10^{19}$ eV and are mainly iron
nuclei, while the dip scenario requires that CRs with energy above 
$\sim 10^{19}$ eV are mostly protons (with no more than $\sim 15\%$ 
contamination of helium) and that the proton dominated extragalactic 
component is important down to energies around $\sim 10^{18}$ eV. 
The differences in the prediction of the chemical composition of CRs
also represent the tool to possibly discriminate between them.

An important aspect of the dip scenario is that it provides a
description of the transition from galactic to extragalactic cosmic
rays which is consistent with the KASCADE observations. For
comparison, the ankle scenario requires that galactic sources should
be able to accelerate cosmic rays up to $\sim 10^{19}$ eV, which
appears rather challenging on the basis of current knowledge of
acceleration processes in galactic sources (see Sect. 2).

In addition to the two scenarios discussed above, there is a third
one, based on the possibility that the chemical composition
at the source is contaminated by nuclei heavier than hydrogen
\cite{par1,par2}. The propagation of these
elements and their fragmentation in the cosmic photon background
determine a rather complex energy dependent chemical composition at
the Earth, which depends somewhat on the assumptions on the injection
spectra and relative abundancies in the sources. In this model the 
transition between galactic and extragalactic cosmic rays takes place
at $\sim 2\times 10^{18}$ eV and is much smoother than for the dip
scenario. The mixed composition model also appears to agree with 
the fact that the galactic spectrum should end at energies
$10^{17}-10^{18}$ eV. However, it requires a rather flat injection 
spectrum ($\alpha\sim 2.4$) and predicts that the chemical composition
in the transition region has a strong iron and helium contamination.

\section{What are the spectrum and anisotropies of UHECRs?}
\label{sec:GZK}

The spectrum of UHECRs is expected to be characterized by the GZK
feature, a flux suppression at energies around $\sim 10^{20}$ eV, due
to photopion production during the propagation of protons on
cosmological scales.   

The theoretical predictions of this part of the spectrum
are extremely uncertain, being dependent on the injection spectrum,
the distribution and spatial density of the sources and the strength
and topology of the intergalactic magnetic field. The search for
this feature has given inconclusive results so far, mainly due to
the very low statistics of detected events. From the statistical
point of view, the most significant data are those collected by
AGASA, HiRes and the Pierre Auger Observatory. AGASA and HiRes, with
comparable exposures, have results which are discrepant in the
highest energy part: the spectrum of AGASA \cite{AGASA} does not show
the GZK suppression, while HiRes spectrum \cite{HIRES} has a
pronounced GZK feature. 
The numerical simulations of Ref. \cite{danny1} showed
however that, given the small number of collected events, the
discrepancy is in fact at $\sim 2-3\sigma$ level, being further
reduced if the offset in the overall normalization of the spectra 
is attributed to a systematic error in the energy determination. 
A systematic error of $\sim 30\%$ would in fact make the experiments
to reasonably agree with each other. It is important to stress that the
simulation of the propagation of UHECRs was carried out in
\cite{danny1} in the case of a truly continuous distribution 
of the sources (no point sources with finite density). It is also
worth pointing out that a recent re-analysis of the AGASA data
resulted in a reduction of the number of events above $10^{20}$ eV
(Teshima, this meeting), which decreases even further the significance
of the alleged AGASA-HiRes discrepancy. 

The most recent measurement of the spectrum by the Pierre Auger
Observatory \cite{Auger} is in closer agreement with
the HiRes results, although again no conclusive evidence for the
absence of the GZK feature can be claimed so far.

If the sources of UHECRs are not {\it diffuse}, namely if they are
astrophysical sources (and not topological defects or cosmological
relics) the directions of arrival of UHECRs are expected to cluster on
small angular scales. These small scale anisotropies (SSA) contain a
large amount of information on the sources, and can be measured by
using the tool of the two point correlation function of the arrival 
directions of the detected events \cite{danny2}. The two 
point correlation function of the AGASA data \cite{AGASA_SSA}, 
when compared with 
the simulated events provides an estimate of the source number 
density of $\sim 10^{-5}\rm Mpc^{-3}$ \cite{danny2}, 
with a very large uncertainty (about one order of magnitude on 
both sides) that results from the limited statistics of events 
above $4\times 10^{19}$ eV, where the analysis should be carried 
out in order to avoid (or limit) the effects of the galactic magnetic
field. 

If one uses the best fit for the source density $n_s \sim 10^{-5}\rm
Mpc^{-3}$ \cite{danny2} and determines the simulated spectrum of 
UHECRs at the Earth, the AGASA small scale anisotropies appear to be
inconsistent with the spectrum measured by the same experiment at the
level of $\sim 5\sigma$ \cite{danny3}.  
Given the large error in the determination of $n_s$ this result cannot 
be taken too seriously, but it certainly hints to the possibility 
that the SSA observed by AGASA may be a statistical fluctuation 
(the HiRes experiment does not find evidence for SSA
\cite{HIRES_SSA}). Its statistical significance has in fact been shown
to be rather weak \cite{finley} and dependent upon the choice of the 
binning angle for the arrival directions \cite{finley,danny3}. The 
HiRes experiment does not find evidence for SSA \cite{HIRES_SSA}.

The combined analysis of the spectrum and SSA is likely to provide us
with precious informations when the full scale Auger data will be
available, although the analysis will be difficult even with Auger
South. In \cite{danny4} the simulations of propagation from point
sources were repeated for the expected Auger South 
statistics of events: the results suggest that it will be
easy to distinguish the case of point sources from the case of a
purely homogeneous distribution, but it will not be easy to achieve a
good resolving power between different values of the source density
$n_s$. The galactic magnetic field and the luminosity function of 
the sources contribute to emphasize this difficulty.
On the other hand, the SSA should also result in the appearance of
{\it hot spots} in the UHECR sky, so that the search for the sources
can proceed though alternative and probably more efficient routes, 
such as the identification of counterparts or the cross-correlation 
of arrival directions with the positions of sources in given 
catalogs. The power of these analyses is expected to be outstanding 
when both Auger South and Auger North will be available. 

\section{Conclusions}
\label{sec:concl}

We discussed three of the open questions in the investigation of the
origin of ultra-high energy cosmic rays: 1) How does cosmic ray
acceleration work? 2) Where does the transition from galactic to 
extragalactic cosmic rays occur? 3) what are the spectrum and 
anisotropies of UHECRs?

In the field of particle acceleration, some of the most impressive
developments are taking place in the investigation of particle
acceleration at shock waves. In particular the theoretical framework
in which the reaction of the accelerated particles on the shocked
plasma is taken into account is getting better defined, and the
amplification of the magnetic field as due to streaming instability is
receiving increasing attention. These studies, together with an
exciting observational situation in which cosmic ray observations join
with X-ray and gamma ray observations in providing us with a more 
consistent picture of particle acceleration at supernova shocks, is
helping us to extend the applicability of the Physical principles
involved to other classes of sources, possibly of relevance for
UHECRs.  

These observations suggest that the galactic component of cosmic rays
ends at energies $\sim 2\times 10^{17}$ eV, therefore hinting to the
fact that the ankle may not be the feature signalling for the
transition from galactic to extragalactic cosmic rays. On the other
hand, observations seem compatible with the {\it dip scenario} and 
with the so-called mixed composition scenario. 
An accurate measurement of the chemical composition between $10^{17}$
eV and $10^{19}$ eV seems to be the most efficient way to discriminate
between these two scenarios of the transition. 

At larger energies, a crucial issue is represented by the
measurement of the spectrum at energies around $10^{20}$ eV, where
the GZK suppression has long been searched and missed. AGASA and HiRes
spectra are in a mild contradiction with each other which may well be
due to statistical fluctuations and a systematic error in the energy 
determination with the two different techniques used for the
measurement. The Pierre Auger Observatory should soon settle this
issue. 

At sufficiently high energies the magnetic field of the Galaxy is
not expected to bend the trajectories of high energy particles by
more than $1-2$ degrees, comparable with the angular resolution of 
the operating cosmic ray experiments. This may lead to the 
identification of small scale anisotropies, flagging the presence 
of discrete sources of UHECRs. A signal of this type was found by 
AGASA, though its statistical significance was later questioned
\cite{finley,danny4}. 
The level of SSA detected by AGASA would correspond to a source density 
$\sim 10^{-5}\rm Mpc^{-3}$ \cite{danny2}. The combination
of spectral analyses and SSA is expected to be fruitful in upcoming
cosmic ray experiments such as the Pierre Auger Observatory, where
additional tools can be adopted for the identification of the sources
(e.g. cross correlation with source catalogs and multifrequency
observations of potential cosmic ray sources). Besides trying to
assess the nature of the sources from statistical analyses, it seems
of the highest priority to attempt to measure the spectrum of a single
source of UHECRs, a finding that would definitely open the era of
UHECR astronomy (see \cite{danny2} for statistical considerations on 
the feasibility of this measurement).

\section*{Acknowledgments}
I am thankful to R. Aloisio, E. Amato, V. Berezinsky, D. De Marco 
and A.V. Olinto for numerous discussions and ongoing collaboration.


\section*{References}

\begin{thebibliography}{99}

\bibitem{gzk1}
Greisen K, 1966 {\it Phys Rev Lett} {\bf 16} 748

\bibitem{gzk2}
Zatsepin G T and Kuzmin V A, 1966 {\it Sov Phys JETP Lett} {\bf 4} 78 

\bibitem{hillas}
Hillas M A, 2005 {\it J Phys G} {\bf 31} 95

\bibitem{wibig}
Wibig T and Wolfendale A W, 2005 {\it J Phys G} {\bf 31} 255

\bibitem{dip1}
Berezinsky V S, Gazizov A and Grigorieva S, 2002 {\it Preprint
hep-ph/0204357}

\bibitem{dip2}
Berezinsky V S, Gazizov A and Grigorieva S, 2005 {\it Phys. Lett} 
{\bf B612} 147

\bibitem{par1}
Allard D, Parizot E, Khan E, Goriely S and Olinto A V, 2005a 
{\it Preprint astro-ph/0505566}.

\bibitem{par2}
Allard D, Parizot E and Olinto A V, 2005b, {\it Preprint
astro-ph/0512345}.

\bibitem{bell78}
Bell A R, 1978 {\it MNRAS} {\bf 182} 443

\bibitem{lagage83a}
Lagage P O and Cesarsky C J, 1983a {\it A\&A} {\bf 118} 223

\bibitem{lagage83b}
Lagage P O and Cesarsky C J, 1983b {\it A\&A} {\bf 125} 249

\bibitem{lb00}
Lucek S G and Bell A R, 2000 {\it MNRAS} {\bf 314} 65

\bibitem{bell2004}
Bell A R, 2004 {\it MNRAS} {\bf 353} 550

\bibitem{malkov}
Malkov M A and Drury L O'C, 2001 {\it Rep Prog Phys} {\bf 64} 429 

\bibitem{amato1}
Amato E and Blasi P, 2005 {\it MNRAS Lett} {\bf 364} 76

\bibitem{amato2}
Amato E and Blasi P, 2006 {\it MNRAS} {\bf 371} 1251

\bibitem{don}
Vladimirov A, Ellison D C, Bykov A, {\it Preprint astro-ph/0606433}

\bibitem{blasi2006}
Blasi P, Amato E and Caprioli D, {\it submitted to MNRAS}

\bibitem{hora}
H\"{o}randel J R, {\it A review of experimental results at the
knee}, in the "Workshop on Physics of the End of the Galactic Cosmic
Ray Spectrum", Aspen, April 25 - 29, 2005 ({\it Preprint
astro-ph/0508014})

\bibitem{warren}
Warren J S et al, 2005 {\it Ap J} {\bf 634} 376

\bibitem{alo}
Aloisio R and Berezinsky V S, 2005 {\it Astrophys. J.} {\bf 625} 249

\bibitem{lemoine}
Lemoine M, 2005 {\it Phys Rev} {\bf D71} 3007

\bibitem{usall}
Aloisio R, Berezinsky V S, Blasi P, Gazizov A, Grigorieva
S and Hnatyk B, 2006 {\it Preprint astro-ph/0608219} ({\it
  Astropart. Phys. in press}

\bibitem{kache}
Kachelriess M and Semikoz D V, 2006 {\it Phys Lett} {\bf B634} 143

\bibitem{danny1}
De Marco D, Blasi P and Olinto A V, 2003 {\it Astropart.Phys}
{\bf 20} 53

\bibitem{Auger}
Sommers P et al. (The Pierre Auger Collaboration), {\it Proceedings
of the 29th International Cosmic Ray Conference}, Pune, India (2005)

\bibitem{AGASA}
Hayashida M et al., 2000 {\it Astron J} {\bf 120} 2190

\bibitem{HIRES}
Abbasi R U et al., 2004 {\it Phys Rev Lett} {\bf 92} 151101

\bibitem{danny2}
Blasi P and De Marco D, 2004 {\it Astropart.Phys} {\bf 20} 559

\bibitem{AGASA_SSA}
Takeda M et al., 1999 {\it Astrophys J} {\bf 522} 225 

\bibitem{danny3}
De Marco D, Blasi P and Olinto A V, 2006a {\it JCAP} {\bf 0601} 002

\bibitem{finley}
Finley C B and Westerhoff S, 2004 {\it Astropart Phys} {\bf 21} 359

\bibitem{HIRES_SSA}
Abbasi R U et al, 2004 {\it Astrophys J Lett} {\bf 610} 73

\bibitem{danny4}
De Marco D, Blasi P and Olinto A V, 2006b {\it JCAP} {\bf 0607} 015

\end{thebibliography}
\end{document}